\documentclass[prl,twocolumn,showpacs,aps]{revtex4}
\usepackage{graphicx,graphics,color,epsfig}
\usepackage{bm}
\usepackage{amsmath}
\usepackage{amssymb}
\usepackage{epstopdf}

\begin{document}

\title{Theory of Mixed-State Effect on Quasiparticle Interference in Iron-based Superconductors}

\author{Yi Gao,$^{1}$ Huaixiang Huang,$^{2}$ and Peiqing Tong$^{1}$}
\affiliation{$^{1}$Department of Physics and Institute of Theoretical Physics,
Nanjing Normal University, Nanjing, Jiangsu, 210046, China\\
$^{2}$Department of Physics, Shanghai University, Shanghai 200444, China}

\begin{abstract}

Based on a phenomenological model with $s_{\pm}$ or $s$-wave pairing
symmetry, the mixed-state effect on quasiparticle interference in iron-based superconductors is investigated by solving
large-scale Bogoliubov-de Gennes equations based on the Chebyshev polynomial expansion. Taking into account the presence of
magnetic field, our result for the $s_{\pm}$ pairing is in qualitative
agreement with recent scanning tunneling microscopy experiment while for the $s$-wave
pairing, the result is in apparent contradiction with experimental observations, thus excluding the
$s$-wave pairing. Furthermore, we treat the effect of vortices rigorously instead of approximating the vortices as magnetic impurities, thus our results are robust and should be more capable of explaining the experimental data.

\end{abstract}


\maketitle

\emph{Introduction}.---The discovery of iron-based superconductors \cite{I1} has
triggered efforts to elucidate the superconducting (SC) pairing mechanism in these materials. One
hotly debated issue is the symmetry and structure of the SC gap.
Theoretically, it was initially suggested that the pairing may be established via spin fluctuations, leading to the so-called
$s_{\pm}$ pairing symmetry ($\Delta_{\mathbf{k}} \sim \cos k_{x}+\cos k_{y}$ defined in the
2Fe/cell Brillouin zone). In this case the SC order parameter (OP) exhibits a sign reversal between the hole pockets (around the $\Gamma$ point) and
electron pockets (around the $M$ point) \cite{I20}. Later, $s$-wave pairing symmetry without sign reversal was also proposed to be a possible candidate which is induced by orbital fluctuations due to the electron-phonon interaction \cite{s++}. Experimentally, the results
about the pairing symmetry remain highly controversial as well. For example,
in Ba$_{0.6}$K$_{0.4}$Fe$_{2}$As$_{2}$, an optimally hole-doped
iron-based superconductor \cite{I3}, the SC gaps measured by the angle
resolved photoemission spectroscopy can be approximately
fitted by $|\Delta_{\mathbf{k}}| \sim |\cos k_{x}+\cos k_{y}|$, with
almost isotropic gaps on all the Fermi surfaces (FS)
\cite{I21-10}, indicating the possible pairing symmetry to be either
$s_{\pm}$ or $s$-wave. Neutron scattering experiments observed a resonance peak at $\textbf{Q}=(\pi,\pi)$ below the SC transition temperature \cite{ns} as predicted by some theoretical works assuming $s_{\pm}$ symmetry \cite{nstheory}, thus at first glance they seemed to support each other. However, later theoretical works suggested that the experimentally observed resonance peak can also be reproduced by assuming $s$-wave pairing \cite{nstheory2}.

Recently, in order to clarify the pairing symmetry, Hanaguri \emph{et al.} used scanning tunneling microscopy (STM) to image the quasiparticle interference (QPI) patterns in the SC state \cite{hanaguri}. They proposed that the relative sign of the SC OP can be determined from the magnetic-field dependence of quasiparticle scattering amplitudes and claimed that their experimental data were only consistent with the $s_{\pm}$ scenario but not the $s$-wave one. Soon after, the experimental results were put into question and argued instead to arise from the Bragg scattering but not due to the QPI because the observed peaks are too sharp \cite{comment}. On the other hand, the magnetic field will induce vortices into the system and lead to the inhomogeneity of the pairing OP in real space, thus affecting the QPI patterns. Theoretical analyzes performed previously have investigated the mixed-state effect on the QPI \cite{ampsur,kpappr,zeeman}. However, in Ref. \cite{ampsur}, only amplitude suppression of the OP near the vortex core was considered, without taking into account the phase variation. In Ref. \cite{kpappr}, the mixed-state effect was treated by using quasiclassical approximation and the QPI derived in this way was rather broad compared to the experimental observation. Furthermore, in Ref. \cite{zeeman}, only the effect of the Zeeman splitting was considered, which should be negligible with respect to the mixed-state effect. Thus there still lacks direct theoretical confirmation of the mixed-state effect on the QPI.

Therefore in this work we adopt a phenomenological model with $s_{\pm}$ pairing symmetry to study the influence of vortices on the QPI by directly solving large-scale Bogoliubov-de Gennes (BdG) equations in real space based on the Chebyshev polynomial expansion. For comparison, the problem is also studied for $s$-wave pairing. In this way the mixed-state effect on the QPI can be rigorously investigated and the results unambiguously support $s_{\pm}$ pairing symmetry in iron-based superconductors.

\emph{Method}.---We begin with an effective two-orbital model on a
two-dimensional lattice \cite{M55}, with a phenomenological form for
the intraorbital pairing terms. The Hamiltonian can be written as
$$H=-\sum_{ij,\alpha\beta,\sigma}[t^{'}_{ij,\alpha\beta}+(\mu-V_{imp}\delta_{ii_{imp}})\delta_{ij}\delta_{\alpha\beta}]
c^{\dag}_{i\alpha\sigma}c_{j\beta\sigma}$$
$$+\sum_{ij,\alpha\beta}(\Delta_{ij,\alpha\beta}c^{\dag}_{i\alpha\uparrow}c^{\dag}_{j\beta\downarrow}+H.c.).\eqno{(1)}$$
Here $i$ and $j$ are the site indices while $\alpha,\beta=1,2$ are the orbital
ones. $\sigma$ represents the spin and $\mu$ is the
chemical potential. Then we consider potential scattering by nonmagnetic impurities through $V_{imp}$ with $i_{imp}$ being the locations of the impurities.
$\Delta_{ij,\alpha\beta}=\frac{V_{ij}\delta_{\alpha\beta}}{2}(\langle
c_{j\beta\downarrow}c_{i\alpha\uparrow}\rangle-\langle
c_{j\beta\uparrow}c_{i\alpha\downarrow}\rangle)$ is the intraorbital
spin singlet bond OP, where $V_{ij}$ is the onsite
[$i=j$] or next-nearest-neighbor [$i=j\pm(\hat{x}\pm\hat{y})$]
attraction which we choose to achieve the $s$-wave or $s_{\pm}$
pairing symmetry, respectively. In the presence of a magnetic field $B$ perpendicular
to the plane, the hopping integral can be expressed as
$t^{'}_{ij,\alpha\beta}=t_{ij,\alpha\beta}$exp$[i\frac{\pi}{\Phi_{0}}\int_{j}^{i}\mathbf{A}(\mathbf{r})\cdot
d\mathbf{r}]$, where $\Phi_{0}=hc/2e$ is the SC flux quantum, and
$\mathbf{A}=(-By,0,0)$ is the vector potential in the Landau gauge.
The hopping integrals are
\begin{equation}
t_{ij,\alpha\beta}=\begin{cases}
t_{1}&\text{$\alpha=\beta,i=j\pm\hat{x}(\hat{y})$},\\
\frac{1+(-1)^{j}}{2}t_{2}+\frac{1-(-1)^{j}}{2}t_{3}&\text{$\alpha=\beta,i=j\pm(\hat{x}+\hat{y})$},\\
\frac{1+(-1)^{j}}{2}t_{3}+\frac{1-(-1)^{j}}{2}t_{2}&\text{$\alpha=\beta,i=j\pm(\hat{x}-\hat{y})$},\\
t_{4}&\text{$\alpha\neq\beta,i=j\pm(\hat{x}\pm\hat{y})$},\\
0&\text{otherwise}\nonumber.
\end{cases}\eqno{(2)}
\end{equation}

The Chebyshev polynomials can be written as $\phi_{k}(x)=\cos[k\arccos x]$ and statisfy
$$\phi_{0}(x)=1, \phi_{1}(x)=x, \phi_{k}(x)=2x\phi_{k-1}(x)-\phi_{k-2}(x),$$
$$\sum_{k=0}^{\infty}\frac{W(x)}{\nu_{k}}\phi_{k}(x)\phi_{k}(x')=\delta(x-x'),\eqno{(3)}$$
where $W(x)=1/\sqrt{1-x^{2}}$, $\nu_{k}=\pi(1+\delta_{k0})/2$ and $x\in[-1,1]$.
Next we define the Green's function matrix:
$$G(\tau)=-\langle T\tau C(\tau)C^{\dag}(0)\rangle,\eqno{(4)}$$
with $C^{\dag}=(\cdots,c^{\dag}_{j1\uparrow},c^{\dag}_{j2\uparrow},\cdots,c_{j1\downarrow},c_{j2\downarrow},\cdots)$.
Eq. (1) can be diagonalized as
$$H=C^{\dag}MC=C^{\dag}QQ^{\dag}MQQ^{\dag}C=\Phi^{\dag}D\Phi.\eqno{(5)}$$
Here $Q$ is a unitary matrix that satisfies $(Q^{\dag}MQ)_{rs}=D_{rs}=\delta_{rs}E_{s}$ and the spectral function can be expressed as \cite{nagai}
$$d_{rs}(\omega)=-\frac{1}{2\pi i}[G_{rs}(\omega+i\eta)-G_{rs}(\omega-i\eta)]$$
$$=\sum_{\gamma}Q_{r\gamma}Q_{s\gamma}^{*}\delta(\omega-E_{\gamma})=\frac{1}{a}\sum_{\gamma}Q_{r\gamma}Q_{s\gamma}^{*}\delta(\tilde{\omega}-\xi_{\gamma})$$
$$=\frac{1}{a}\sum_{\gamma}Q_{r\gamma}Q_{s\gamma}^{*}\sum_{k=0}^{\infty}\frac{W(\tilde{\omega})}{\nu_{k}}\phi_{k}(\tilde{\omega})\phi_{k}(\xi_{\gamma})$$
$$=\frac{1}{a}\sum_{k=0}^{\infty}\frac{W(\tilde{\omega})}{\nu_{k}}\phi_{k}(\tilde{\omega})\phi_{k}(\tilde{M})_{rs},$$
$$d(\omega)=\frac{1}{a}\sum_{k=0}^{\infty}\frac{W(\tilde{\omega})}{\nu_{k}}\phi_{k}(\tilde{\omega})\phi_{k}(\tilde{M}).\eqno{(6)}$$
Here $r$, $s$, $\gamma=1,\cdots,L$ with $L=4N_{x}N_{y}$ and $N_{x}$ ($N_{y}$) being the number of
lattice sites along $\hat{x}$ ($\hat{y}$) direction of the 2D
lattice. $a=(E_{\gamma}^{max}-E_{\gamma}^{min})/(2-\epsilon)$ ($\epsilon>0$ is a small number), $b=(E_{\gamma}^{max}+E_{\gamma}^{min})/2$, $\tilde{\omega}=(\omega-b)/a$, $\xi_{\gamma}=(E_{\gamma}-b)/a$ and $\tilde{M}=(M-bI)/a$.

If we further define the L-dimensional vectors $e(o)$ and $h(o)$ as
$e(o)_{\gamma}=\delta_{\gamma o}$ and $h(o)_{\gamma}=\delta_{\gamma o+2N_{s}}$ ($N_{s}=N_{x}N_{y}$ and $o=1,\cdots,2N_{s}$), we can express the self-consistent parameters as

$$n_{j\beta\uparrow}=\langle c_{j\beta\uparrow}^{\dag}c_{j\beta\uparrow}\rangle=\int_{-\infty}^{\infty}d\omega f(\omega)d_{mm}(\omega)$$
$$=\int_{-\infty}^{\infty}d\omega f(\omega)e(m)^{T}d(\omega)e(m)$$
$$=\sum_{k=0}^{N-1}g_{k}e(m)^{T}e_{k}(m)\frac{T_{k}}{\nu_{k}},$$

$$n_{j\beta\downarrow}=1-\sum_{k=0}^{N-1}g_{k}h(m)^{T}h_{k}(m)\frac{T_{k}}{\nu_{k}},$$

$$\Delta_{ij,\alpha\beta}=\frac{V_{ij}\delta_{\alpha\beta}}{2}\sum_{k=0}^{N-1}g_{k}[e(n)^{T}h_{k}(m)+e(m)^{T}h_{k}(n)]\frac{T_{k}}{\nu_{k}},\eqno{(7)}$$
where $m=2(j_{y}+N_{y}j_{x})+\beta$ and $n=2(i_{y}+N_{y}i_{x})+\alpha$ with $i_{x},j_{x}=0,\cdots,N_{x}-1$ and $i_{y},j_{y}=0,\cdots,N_{y}-1$. $e_{k}(m)=\phi_{k}(\tilde{M})e(m)$, $h_{k}(m)=\phi_{k}(\tilde{M})h(m)$ and $T_{k}=\int_{-1}^{1}d\tilde{\omega}f(a\tilde{\omega}+b)W(\tilde{\omega})\phi_{k}(\tilde{\omega})$. $N$ is the cutoff in the summation and $g_{k}$ is the kernel we convolute to avoid Gibbs oscillations \cite{kernel}. In addition, $f(x)$ is the Fermi distribution function. At zero temperature we have
\begin{equation}
T_{k}=\begin{cases}
\pi-\arccos(-\frac{b}{a})&\text{$k=0$},\\
\\
-\frac{\sin[k\arccos(-\frac{b}{a})]}{k}&\text{$k\neq0$}\nonumber.
\end{cases}\eqno{(8)}
\end{equation}
Then we can solve the BdG equations self-consistently and the chemical potential is determined by the doping concentration. The calculation is repeated until the absolute error of
the OP between two consecutive iteration steps and that of the total electron number
are less than $10^{-4}$. The local density of states (LDOS) is given by
$$\rho_{j}(\omega)=\frac{1}{a}\sum_{\beta}\sum_{k=0}^{N-1}\frac{g_{k}}{\nu_{k}}[W(\tilde{\omega})\phi_{k}(\tilde{\omega})e(m)^{T}e_{k}(m)$$
$$+W(\tilde{\omega'})\phi_{k}(\tilde{\omega'})h(m)^{T}h_{k}(m)],\eqno{(9)}$$
where $\tilde{\omega'}=(-\omega-b)/a$.

The benefits of this method are threefold. First, it requires much less storage than the exact diagonalization method since the matrix $M$ is sparse, thus we can solve large-scale BdG equations and obtain the QPI by Fourier transforming the real-space LDOS in sufficiently wide range. Second, it is applicable in parallel computation because the self-consistent parameters on each lattice site can be calculated separately. Third, the expansion scheme is very stable and efficient.

In our calculation, the magnitudes of the parameters are chosen as $t_{1-4}=1,0.4,-2,0.04$.
Magnetic unit cells are introduced where each unit cell accommodates
four SC flux quanta and the linear dimension is $N_{x}\times
N_{y}=80\times80$, corresponding to a magnetic field $B\approx8.32$
Tesla, close to the experimental value ($10$ Tesla) \cite{hanaguri}. $V_{ii}$ and $V_{ij}$ [$i=j\pm(\hat{x}\pm\hat{y})$] are
chosen to be $-2.8$ and $-2$, respectively. Moreover we introduce 12 randomly distributed impurities with $V_{imp}=0.3$. $E_{\gamma}^{max}$ ($E_{\gamma}^{min}$) is chosen as $1.5$ ($-1.5$) band width. Throughout the paper, we
set the system to be $20\%$ hole-doped. In calculating the self-consistent parameters, we use the Jackson kernel
$$g_{k}=\frac{(N-k+1)\cos\frac{\pi k}{N+1}+\sin\frac{\pi k}{N+1}\cot\frac{\pi}{N+1}}{N+1},\eqno{(10)}$$
with $\epsilon=0.001$ and $N=500$. For the LDOS we convolute the Lorentz kernel
$$g_{k}=\frac{\sinh[\lambda(1-\frac{k}{N})]}{\sinh\lambda},\eqno{(11)}$$
with $\lambda=4$, $\epsilon=0.004$ and $N=\lambda/\epsilon$.

\begin{figure}
\includegraphics[width=1\linewidth]{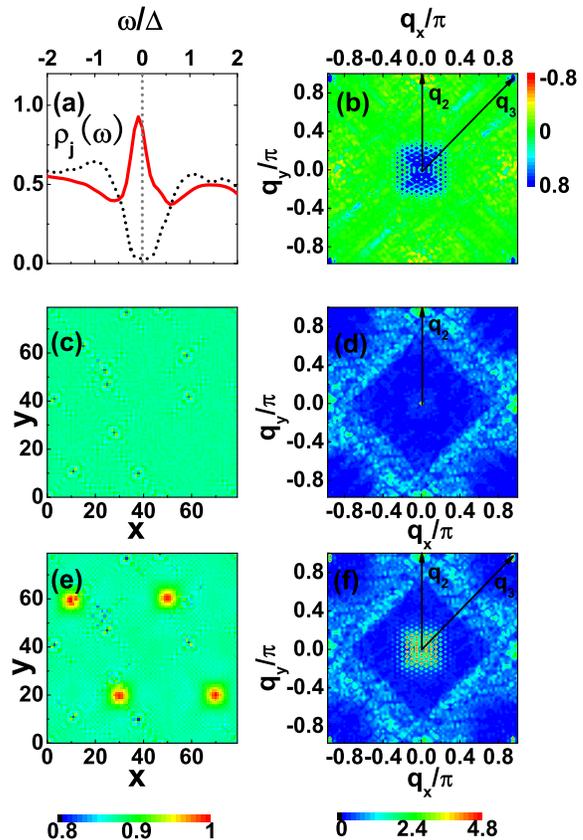}
 \caption{\label{fig1} (color online) (a) The LDOS at the vortex core center (red solid) and at $B=0$ (black dot) as a function of the reduced energy $\omega/\Delta$. Here $2\Delta$ is the SC gap between two SC coherence peaks in the LDOS at $B=0$. The gray-dotted line indicates the position of $\omega=0$. (b) $Z_{\mathbf{q}}(B\neq0)-Z_{\mathbf{q}}(B=0)$ at $\omega=\Delta$. (c) and (d) show $Z_{j}(\omega=\Delta)$ and $Z_{\mathbf{q}}(\omega=\Delta)$ at $B=0$, respectively. (e) and (f) are the same as (c) and (d), respectively, but for $B\neq0$. (c) and (e) share the same color scale while the case is similar for (d) and (f).}
\end{figure}
\emph{Results and discussion}.---First we consider the $s_{\pm}$
case. For $V_{imp}=0$, Fig. \ref{fig1}(a) shows that there exists a negative-energy in-gap peak in the LDOS at the vortex core center, in agreement with the experimental observation \cite{stm} and our previous results based on the exact diagonalization method \cite{gao}, indicating the reliability of the Chebyshev polynomial-expansion scheme. Figure \ref{fig1}(c) plots the spatial distribution of $Z_{j}(\omega)=\rho_{j}(\omega)/\rho_{j}(-\omega)$ as defined in Ref. \cite{hanaguri}, for $B=0$ and at $\omega=\Delta$. The locations of the impurities can be clearly identified as the low-intensity spots. The corresponding $Z_{\mathbf{q}}(\omega)$, which is the Fourier transformation of $Z_{j}(\omega)$, is shown in Fig. \ref{fig1}(d). We notice that there are high-intensity peaks at $\mathbf{q}_{2}$ [$(\pm\pi,0)$ and $(0,\pm\pi)$] which arise from the interpocket scattering between the hole and electron pockets. Furthermore the intensities at $\mathbf{q}_{3}$ [$(\pm\pi,\pm\pi)$] are much weaker than those at $\mathbf{q}_{2}$, consistent with the experimental observation \cite{hanaguri}. Upon applying the magnetic field, vortices are introduced into the system and their locations are denoted as the high-intensity spots in Fig. \ref{fig1}(e). In this case, from Fig. \ref{fig1}(f) we can see that there exist additional peaks at $\mathbf{q}_{3}$, whose intensities are enhanced by the application of the magnetic field and they are due to the interpocket scattering between different electron pockets (see Fig. 1 in Ref. \cite{hanaguri}). Figure \ref{fig1}(b) shows the magnetic field-induced change in the QPI intensities defined as $Z_{\mathbf{q}}(B\neq0)-Z_{\mathbf{q}}(B=0)$. In the presence of the time-reversal symmetry breaking due to the magnetic field, the sign-preserving scatterings at $\mathbf{q}_{3}$ ($\mathbf{q}_{3}$ connects the FS with the same sign of the SC OP) are enhanced while the sign-reversing scatterings at $\mathbf{q}_{2}$ ($\mathbf{q}_{2}$ connects the FS with the opposite sign of the SC OP) are suppressed. Both the locations and the sharpness of the QPI peaks shown in Figs. \ref{fig1}(b), \ref{fig1}(d) and \ref{fig1}(f) are in reasonable agreement with the experimental data \cite{hanaguri}, suggesting that the experimentally observed peaks are indeed due to the QPI but not the Bragg scattering as argued in Ref. \cite{comment}.

\begin{figure}
\includegraphics[width=1\linewidth]{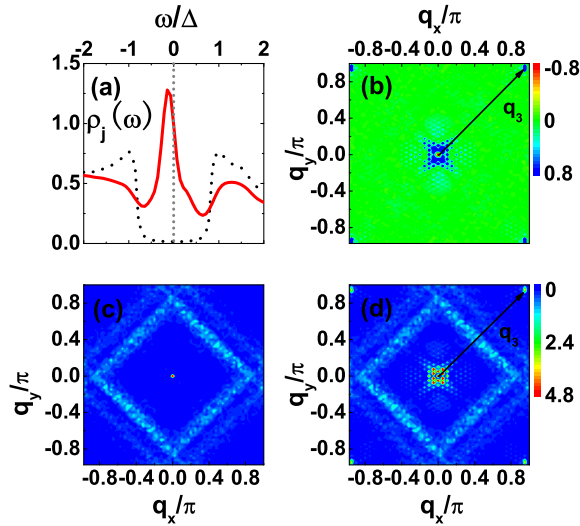}
 \caption{\label{fig2} (color online) (a), (b), (c) and (d) are similar to Figs. \ref{fig1}(a), \ref{fig1}(b), \ref{fig1}(d) and \ref{fig1}(f), respectively. (c) and (d) share the same color scale.}
\end{figure}

Next we consider the $s$-wave case. For $V_{imp}=0$, the LDOS at $B=0$ (black dot) and at the vortex core center (red solid) shown in Fig. \ref{fig2}(a) are also consistent with those obtained by exact diagonalization \cite{gao2}, again suggesting the validity of the current polynomial-expansion scheme. From Fig. \ref{fig2}(c) we notice, the QPI in the presence of impurities at $B=0$ exhibits no pronounced peaks at either $\mathbf{q}_{2}$ or $\mathbf{q}_{3}$, as compared to the clear peaks at $\mathbf{q}_{2}$ in the $s_{\pm}$ case as shown in Fig. \ref{fig1}(d). After applying the magnetic field, the intensities at $\mathbf{q}_{3}$ are enhanced and they form sharp peaks as shown in Fig. \ref{fig2}(d), similar to the $s_{\pm}$ case. At last, from the magnetic field-induced change in the QPI intensities plotted in Fig. \ref{fig2}(b) we can see, the intensities are enhanced at $\mathbf{q}_{3}$ but remain almost unchanged at $\mathbf{q}_{2}$. The lack of distinct structures at $\mathbf{q}_{2}$ is in stark contrast to the $s_{\pm}$ case and is inconsistent with the experimental observations \cite{hanaguri}. Therefore, the different behavior of
the QPI intensities at $\mathbf{q}_{2}$ in the $s_{\pm}$ and $s$-wave pairing cases
makes it possible to distinguish these two types of pairing symmetry
since the STM experiment observed clear structures at $\mathbf{q}_{2}$, thus excluding the possibility of $s$-wave
pairing in iron-based superconductors.

\emph{Summary}.---In summary, by using the Chebyshev polynomial expansion to directly solve large-scale BdG equations in real space, we have investigated the mixed-state effect on QPI in iron-based superconductors by assuming $s_{\pm}$ or $s$-wave pairing symmetry. For the $s_{\pm}$
pairing, the QPI intensities at $\mathbf{q}_{2}$ which connects the FS with the opposite sign of the SC OP are suppressed by the application of the magnetic field while the situation at $\mathbf{q}_{3}$ is reversed where $\mathbf{q}_{3}$ connects the FS with the same sign of the SC OP. The obtained results at both $B=0$ and $B\neq0$ are in qualitative agreement with experiment, suggesting that the experimentally observed peaks are indeed due to QPI. On the other hand, for the $s$-wave pairing, the QPI intensities at $\mathbf{q}_{2}$ are featureless both with and without the magnetic field. Based on the available experimental data, the $s$-wave pairing can be excluded in iron-based superconductors.

{\it Acknowledgments} We thank A. Li, D. G. Zhang, T. Zhou, C. S. Ting and Y. Xiong for helpful discussions. This work was supported by the National Natural Science Foundation of China (Grants No. 10974097 and No. 11175087), and by National Key Projects for Basic Research of China (Grant No. 2009CB929501).

\end{document}